# A Field Study of On-Calendar Visualizations

Dandan Huang[†]  Melanie Tory[‡]  Lyn Bartram[*]
University of Victoria   Tableau Research   Simon Fraser University

**ABSTRACT**

Feedback tools help people to monitor information about themselves to improve their health, sustainability practices, or personal well-being. Yet reasoning about personal data (e.g., pedometer counts, blood pressure readings, or home electricity consumption) to gain a deep understanding of your current practices and how to change can be challenging with the data alone. We integrate quantitative feedback data within a personal digital calendar; this approach aims to make the feedback data readily accessible and more comprehensible. We report on an eight-week field study of an on-calendar visualization tool. Results showed that a personal calendar can provide rich context for people to reason about their feedback data. The on-calendar visualization enabled people to quickly identify and reason about regular patterns and anomalies. Based on our results, we also derived a model of the behavior feedback process that extends existing technology adoption models. With that, we reflected on potential barriers for the ongoing use of feedback tools.

**Keywords**: Personal visualization, feedback design, physical activities, field study, digital calendar.

**Index Terms**: H.5.2. Information interfaces and presentation (e.g., HCI): User interfaces

## 1 INTRODUCTION

People have access to increasing quantities of personal data, ranging from information about their home energy conservation, to their health, to their personal fitness. However, reflecting on these data to make better lifestyle choices often requires substantial effort in data organization / integration / interpretation, a substantial barrier to ongoing adoption of personal informatics (PI) tools [3, 4]. How then, can we engage ongoing use of PI tools in nonprofessional life? We propose that the answer is threefold. First, we should make the information fit naturally into people's life patterns. Second, the tools need to provide sufficient contextual information to explain both the data patterns and their causes. Finally, the effort required to gather and integrate the data should be minimized: simply adding "yet another app" defeats this goal. Towards these ends, we are exploring the idea of integrating such feedback data (personal fitness data in this study) into people's existing information eco-systems (specifically, their digital calendars). In doing so, we aim to make behavior feedback data readily accessible and comprehensible, achieving the goals of perceived usefulness and ease of use [2].

Feedback tools typically focus on three types of questions: *what*

[†] email: dhuang@uvic.ca
[‡] email: mtory@tableau.com
*email: lyn@sfu.ca

("what is the current status or progress"), *why* ("why are the data patterns like this", "why did an anomaly happen"), and *how* ("how could I improve"). The commonly used *persuasive* design strategy usually focuses on *how*, promoting action in the moment. In contrast, we examine how visualization design could enhance the *reflective* understanding of one's behaviors. Our general philosophy centers around helping people understand their behavior with operational understanding [3] rather than lecturing them into behavior change. We view understanding as a first (but not the only) step towards encouraging new habits; therefore, tools that can succeed at this step have made progress.

Existing feedback tools are typically dedicated apps and web portals, with limited mash-up capabilities to combine with rich contextual data. Thus they are typically very good at providing the raw data (what), but are less effective at helping people reason about underlying causes (why), often due to lack of support for revealing temporal patterns and related personal activities [4]. Lack of context for reflection has been reported as one of the major barriers for personal information systems [14]. For example, a recent study [18] showed that it is very difficult for people to link their daily routines to their residential energy consumption, making it difficult to take meaningful action. Thus, finding the appropriate contextual framing is a critical factor in helping people recognize and understand information patterns [4].

We aim to design tools that provide daily context to better understand one's feedback data, and meanwhile are easy to access and manage with minimal dedicated effort. As a first step in this direction, we propose to integrate a visualization of personal data (fitness data in the current study) into a digital calendar. We expect that incorporating data feeds into one's existing information tools should reduce learning effort, reduce the time and dedicated effort required to access the data and assemble related contextual information, and ensure that the data is encountered on a regular basis to support awareness.

In the current study we are particularly interested in the role of a digital calendar for providing context: can calendar events help people to interpret and understand their feedback data? To explore this design concept, we implemented an on-calendar visualization linked to Google Calendar and investigated its value for displaying fitness feedback (from Fitbit) in an eight-week field study. Our analysis focuses primarily on two questions: RQ1: *To what extent can people use calendar data as context for reasoning about their fitness data?* And RQ2: *How do people react to the idea of integrating feedback data into their personal calendars?*

Our results confirm that a personal calendar can provide rich context for people to reason about patterns and anomalies in their behavior feedback data. Participants liked the idea of integrating personal data within a calendar visualization and found it non-disruptive and easy to access and understand. Based on our results, we propose a model of the feedback process that adapts and extends existing technology adoption models. It offers a starting point that designers can use to reason about the role of feedback tools and potential barriers to ongoing adoption. Our research demonstrates the importance of integrating feedback data with relevant contextual information to support reasoning.

## 2 RELATED WORK

We focus on related work in three key areas: persuasive and reflective strategies for feedback design, personal visualization, and user studies of feedback tools.

### 2.1 Design Strategies for Feedback Tools

Visualization feedback tools have been developed for a variety of applications, especially for monitoring energy conservation [7, 8, 11] and physical activity [6, 12]. For an overview of this work, see Froehlich et al.'s survey of eco-feedback design [10].

In many feedback designs, persuasion is the dominant approach. Persuasive technologies typically encourage people to take "expected" actions. However, they usually do not provide much background reasoning to help people understand their behavior choices and consequences [1]. As such, the persuasive approach has been criticized [20, 22]. Strengers et al. questioned the value of in-home display approaches [20] for energy conservation. They found that household behavior change cannot be modeled by one or two variables; instead, it is mediated in the context of everyday life, socially, culturally and institutionally. Critics of persuasive technology suggest a shift from prescription to reflection. For example, the Dubuque system engages people to reflect about resource consumption [8]. A study showed that the reflection-oriented design helped participants increase their understanding of water consumption and also encouraged social conversation about water conservation. Our design approach follows a similar reflection-oriented philosophy.

### 2.2 Personal Visualization

Personal visualization research explores the unique design and evaluation challenges in designing visualizations for use in personal contexts [14]. One key research and design challenge for personal visualizations is how to provide personal context data that may be beyond what sensor technology could capture, e.g., social and cultural background, personal experience, skill sets, etc. Our work explores a design direction for incorporating one type of contextual information (calendar information).

Meanwhile, personal visualizations are expected to fit easily into people's everyday life routines, minimizing the effort for on-going use. This complies with "ease of use", one key in the Technology Acceptance Model (TAM) [2]. Temporal drop-offs in use are quite common with feedback systems [4]. In the field study with Dubuque [8], 40% of the participants reported that they rarely used the system. We conjecture that having to consciously access and login to a dedicated application may have been a strong contributing factor. We propose that integrating data visualization into a regularly used tool might engage people in ongoing monitoring.

### 2.3 Evaluating Feedback Applications

Feedback applications have been studied extensively, most often in field studies. For example, Consolvo et al. explored design requirements by deploying a mobile sensing system (Houston) in the field [5]. Fish'n'Steps [16] investigated social influence related to feedback use. Most of these studies focused on measuring behavior change influenced by interventions. However, measuring behavior change might be inaccurate and unnecessary [15]. In contrast, our focus is on understanding how easy access to contextual information can help people to understand their behavior data. Most similar to our work is a long-term persuasive technology study of Fitbit [9]. However, unlike their work, our primary interest is to explore a non-persuasive design approach and see how people react to the idea of integrating feedback data on their personal calendar. Meanwhile, different from randomized controlled trials commonly used in health informatics research, our goal is to conduct an early phase qualitative investigation of our design approach with open-ended questions.

## 3 VISUALIZATION DESIGN AND IMPLEMENTATION

### 3.1 Visualization Design

We carefully considered the visual design used to integrate feedback data into the calendar. To make personal visualization applications fit into everyday life routines, we need to support varying attentional demand; that is, as an additional visualization layer, the data view needs to be subtle enough that it does not interfere with regular use of the digital calendar while remaining perceivable. We conducted a preliminary lab study to identify visual encodings that were perceivable but not disruptive, and selected the best among these for this field deployment.

### 3.2 Implementation and Pilot Studies

We implemented an interactive web application (Figure 1), which worked as an independent online digital calendar, synchronizing with calendar events (through Google API) and also fetching personal data feeds (e.g., Fitbit API). An additional visualization layer showing the personal data stream sat on the calendar background and could be customized. First, the visualization could be displayed either side-by-side or overlapped with calendar events. Users could also choose the visual encoding: either line graph (the default, as in Figure 1) or luminance. To balance the ambience of foreground calendar events and the background data stream, users could adjust settings of the visualization layer, for example, transparency, data stream color (with grey as default) or scale. Additional screenshots illustrating these various settings are available in a supplementary file. Prior to the current field study, we deployed an earlier version of our application in two pilot studies: home energy conservation (with home power meter data) and personal physical activities (with data from Fitbit). Our final design incorporated feedback from these pilots.

## 4 STUDY DESIGN

To answer our two research questions (*To what extent can people use calendar data as context for reasoning about their fitness data? How do people react to the idea of integrating feedback data into their personal calendars?*), we deployed our prototype calendar in an eight-week field study of Fitbit users. Our study compared Visualization and Control groups (who used our calendar prototype and Fitbit's standard feedback tools respectively), enabling us to investigate the influence of providing extra context for reasoning. Our emphasis was on exploring people's experiences with the on-calendar visualizations rather than measuring differences in behavior change. Therefore, we employed a qualitative approach with open-ended research questions rather than statistical comparisons between the groups.

### 4.1 Participants

We recruited participants among existing Fitbit users on our university campus and through social networks. We chose to recruit existing users rather than new users because they already had some motivation to use feedback tools, they were already experienced with Fitbit's basic feedback applications, and for them, using a fitness tracker and its software would not itself be a novelty. We also required that participants be familiar with digital calendars and have a Google account (necessary to use our prototype). In total, we recruited 21 participants with age ranging from 20 to 60+, 14 female and 7 male. Two of them (one female and one male) dropped out after the first two weeks.

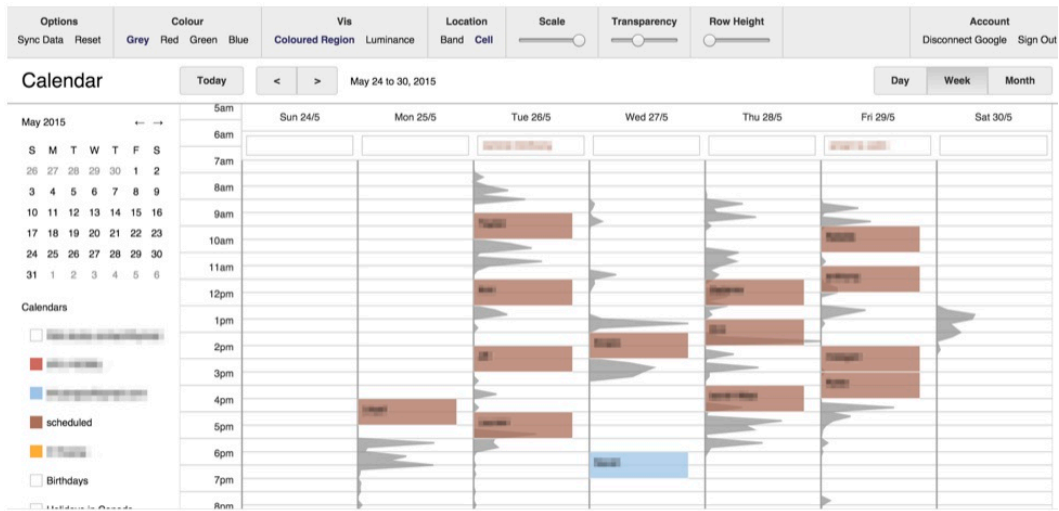

Figure 1. Calendar-based feedback tool used in our study (week view with Fitbit data displayed as a line graph overlapped with calendar events). See more screenshots in the supplementary material.

### 4.2 Conditions

Participants were randomly divided into two groups: **C**ontrol (C1~C9) and **V**isualization (V1~V10). This design allowed us to investigate whether extra context from a personal calendar could improve people's understanding of their feedback data. Participants in the control group used their baseline feedback application (i.e., the Fitbit.com site – see screenshots in the supplementary material). Participants in the Visualization group used the baseline feedback application in the first two weeks; they were then introduced to our web-based calendar visualization after week 2. Visualization group participants were asked to use our calendar application as their primary scheduling and feedback tool; however, they were not prevented from also using their default calendar service (e.g., Google calendar or iCal) and Fitbit's feedback tools.

### 4.3 Procedure

Before the first week, we met participants and introduced the procedure. During the first two weeks we collected baseline information and participants were told to continue using Fitbit as they had done in the past. We interviewed all participants in week 3, during which Visualization group participants were introduced to our on-calendar visualization. To investigate their initial experience and help the participants on technical issues, we interviewed all participants again in week 5. Final interviews were scheduled in week 9. Participants were asked to fill in a weekly International Physical Activity Questionnaire (IPAQ) [23] through an online portal. Reminder emails with the survey link were sent to them on Friday afternoon every week. At the end of the final interview, participants in the control group were also introduced to the on-calendar application and asked for comments.

### 4.4 Data collection

We collected data with weekly surveys, application logs and interviews. We also had access to participants' Fitbit data, but chose not to use this data to measure physical activity level because of its incompleteness (Fitbit cannot accurately capture activities such as cycling, spin class, and swimming). Instead we estimated physical activity (PA) using the weekly IPAQ survey. We dropped V8's survey data from the analysis because only 3 surveys were submitted. The remaining 18 participants submitted at least 6 entries of the online survey. Metabolic Equivalent (MET) is a commonly used physiological measure to assess physical activities. METs of the weekly surveys were calculated according to the scoring protocol of IPAQ [23]. Meanwhile, interactions of participants while using the calendar visualization (e.g., changing the visual encoding or layout) were automatically logged. In the interview, we asked participants to recall their PA patterns, their experience using the feedback tools, and the impact in their life. During the interview, they were also asked to bring up their feedback application and reason about their own data patterns. We observed how they interacted with the application and how they performed tasks to reason about their data.

Our analysis focused on qualitative feedback about the on-calendar design approach. We were most interested in how the approach would influence people's ability to reason about their feedback data, and to what extent they would find the on-calendar visualizations helpful and/or disruptive. We therefore employed a primarily qualitative analysis approach.

## 5 FINDINGS

### 5.1 Physical activity levels

We first examined the physical activity (PA) variation before and after the calendar intervention. The results showed the two groups were not significantly different in MET measures ($t(16)=0.53$, $p=0.60$, Cohen's $d=0.27$). PA tended to increase more for the Visualization group than for the Control group, but this was overshadowed by individual differences (Figure 2: top). Participant comments suggested that behavior change (PA variation) was most influenced by other aspects in their lives, e.g., traveling (V2, V6, C5), relocation (V10, C6), facility service interruption (C3), or a training program (V5, C4). These results are not unexpected: behavior impact of a single intervention can be difficult to quantify in fitness studies. Thus measuring behaviour change was not our main goal. Instead, we focused the majority of our analysis on system use, its role in the feedback process, and how it influenced people's reasoning (next sections).

### 5.2 System Use

Application logs showed 152 visits (user sessions) and 208 user interactions (setting and view changes) during the study. Figure 2 (bottom) shows application use throughout the 8 weeks. The peak

15

usage was in the morning (~10am) and in the evening (~9pm). The application remained active for durations ranging from one minute to four days: some participants brought the application up for a quick look while others continually kept the tab open.

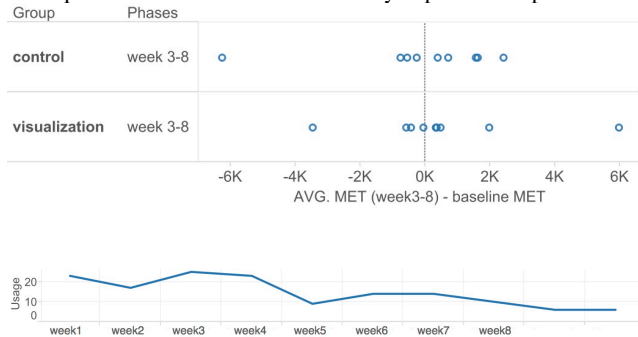

Figure 2. (Top) Change in MET values from weeks 1-2 (baseline) to weeks 3-8 (intervention) for individuals in control and visualization groups. Each mark represents one participant's change in average MET scores. (Bottom) Total time progression of application usage.

Application logs (Figure 3) showed that all participants preferred Colored Region (line graphs) as the visualization setting. They reported that luminance as the visual encoding required extra cognitive effort to understand, and the color made the calendar look busy and interfered with calendar events (particularly when the calendar events were color coded). Grey color and overlapped display were used most often, suggesting that they were least disruptive. Only one participant chose to show the visualization layer in a separate band side-by-side with calendar events. We do note that the favorite settings were also defaults, so it is possible that the defaults influenced participants' preferences. However, we tried to mitigate the effects of defaults by asking participants to explore all options when the application was introduced in the first meeting; the application was implemented to remember the customized settings so the defaults would not appear again.

In the interviews, participants reported that the visualization layer did not interfere with their use of calendar events (V1, V2, V9, V10), especially with the grey color. This suggests that with proper visual encoding, displaying data as an additional layer on a calendar need not interfere with regular calendar use. Most participants stayed on week view most of the time, and switched between week and month views (175 view switches were logged among 10 participants) when they explored data patterns with different time ranges and levels of detail.

### 5.3 A Model of the Behavior Feedback Process

We transcribed the interview recordings and conducted content analysis [17]. The coding process was facilitated by AQUAD (version: 7.4.1.2) [24]. First, the transcripts were open coded with a focus on how feedback tools influence understanding and reasoning about physical activity, what context the participants used for reasoning, interaction with visualization tools, how this understanding relates to one's goals and to changes in behavior, and barriers of current feedback use. Then those codes were clustered and organized into categories of *state* (current physical activity status), *goal* (personal objectives for using feedback tools), *reasoning* (how one makes sense of data patterns), *insights and awareness* (people's understanding of their PA), *behavior choice* (choices about when and how to engage in physical activity) and *emotion* (what emotion could be evoked in the process). We then used the data to build an understanding of relationships between these concepts. This analysis resulted in the behavior feedback model illustrated in Figure 4.

**State** represents data reflecting the current status that is collected and visualized with feedback tools; for example, the current activity level, progress during the day or the week, PA patterns and change. Participants reported various data about state that they would read from feedback tools, including immediate measures in the moment (e.g., active minutes, heart rate, steps) and reflective progress measures (e.g., long-term trend, activity performance, calorie balance, daily and weekly progress towards goals, and sleep quality). Participants used both data summaries and detail views to access this information.

Personal **goals** could be short-term or long-term. As an example of a long-term goal, V7 was using Fitbit feedback to motivate himself to build regular gym routines that could fit in his current schedule. On the other hand, V6 used the feedback tool to track short-term daily and weekly step goals. Personal goals strongly influenced what participants expected to see about their state. V2 had to manage a health condition, so he focused most on sleep quality and resting heart rate. C8, who already had a regular exercise routine, mostly used feedback tools to track her exercise plan (two runs and two gym visits per week). In some cases, goals also influenced data collection. V9, hoping to know the impact of depression on his productivity, set up a daily self-report system to track his mood. V8 replaced her Fitbit device with a different model because she wanted to monitor her cardio status while exercising, a feature that was not possible with the first model.

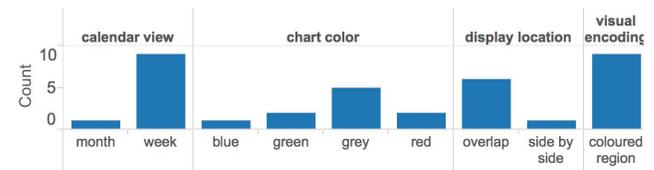

Figure 3. Preferred visualization settings (Visualization group). The most popular visual encoding was a grey line chart overlapped with the calendar data in week view, as shown in Figure 1.

Goals varied widely across our participants. Examples included progress checking (checking daily or weekly progress), in-the-moment monitoring (monitoring heart rate in cardio zone), exploration (exploring what exercise fits better), problem investigation (investigating sleep quality), or medical/physical condition management (managing diabetes). One's goal may vary with age as well. For example, an older participant stated, *"Fat burn, you can get how often I am doing, hitting the cardio level … If I was younger that might be important…I think that probably for older people using the Fitbit, that probably the most important tool is to see that the improvement is there on a daily basis."* (V6)

Personal goals motivate people to look at their data to gain **awareness**, and to **reason** about their data to gain **insights**, by posing and answering questions. We categorized three types of questions: (1) *What* ("What is the current status or performance?", "Do the data accurately reflect my situation?", "Have I done 3 runs this week?", "What are the data patterns in a year/month/week/day?"), (2) *Why* ("Why do I have a trend like this?", "Why is the pattern on Friday night different?", "Why do I always see a spike in my data early in the morning?"), and (3) *How* ("How can I improve?", "How can I fit running into this week's schedule?", "How can I customize my exercise goals?").

People performed a variety of tasks to seek answers to these questions, including: making comparisons, mapping Fitbit data to calendar events, integrating data, looking up items, changing the timeline scale, counting, identifying patterns and anomalies, observing overall trends, searching related domain knowledge, and exploring what-if experiments. During the interviews, participants were asked to reason about their Fitbit data, e.g., peak values, patterns, and anomalies. Tasks they performed most often were *comparisons* (19/19 participants) and *activity mapping* (19/19 participants), in which people related data patterns to activities that happened at the same time. Participants also *compared their progress with goals* (18/19 participants), *baselines* (3/19 participants), *historical performance* (3/19 participants), or *others* (8/19 participants). By relating their activities to Fitbit data, participants could *identify anomalies* (10/10 in Visualization group and 5/9 in Control group) and *reason about regular patterns* (5/10 in Visualization group and 3/9 in Control group). V9 expected to be able to do *what-if exploration* (specifically, adjust his bed time to investigate sleep quality), but none of the feedback tools he used supported this task.

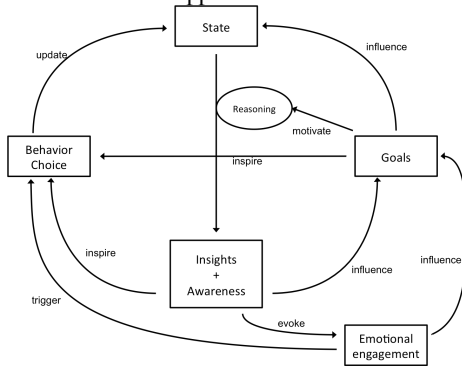

Figure 4. Model of the behavior feedback process

**Insights** derived from **reasoning** helped people to optimize their goals and inspired them to reflect on **behavior choices**. C9 customized her goal based on insights from the historical data, *"I found 9k is reasonable for me to make it every day…but I wanna make the goal everyday for a while before I upgrade it."* (C9) C4 found that keeping the same goal every day was not realistic for her, because it did not take into account the ups and downs of her energy level during the week. Most commonly, insights encouraged immediate action. Many participants reported they took an extra walk to meet their daily step goal (V3, V5, V6, C1, C4, C6, C7, C9). *"I will find this silly to confess, if my steps are like 7,000, 8,000 at the end of the day, I'll jog in front of the television."* (V3) Insights also encouraged participants to adjust their exercise plan according to their progress and their schedule (V2, C3, C5, C8, C9). *"So say I have done two gyms this week, check. And I have two runs this week, check. If I just did one, I should find some time to fix it."* (C8) Other participants identified motivational strategies that might work for them by reflecting on the historical data. *"….because if I schedule it, like, I don't think an hour -- it doesn't really take an hour, it takes an hour to do the workout, it takes ten minutes in the change room and then 20 minutes to get there, 20 minutes to get, like I need to have that full time, and so otherwise I schedule things too close."* (C4) The analysis also helped people select different types of exercises. *"I don't know why, it's easy to fit in a walk or a run but with the same amount of time I won't do weights or push ups."* (C7).

We also observed that reasoning about data could evoke **emotional engagement**. C8 found that her sleep data reflected how her son was doing over night. V6 could view the step data of her sister from the challenge feature (on the Fitbit application), and once identified some anomalies in her sister's data: it turned out her sister was sick. More common examples included the enjoyment of reminiscing and the satisfaction of keeping to an exercise plan (V3, V6, V9, C4, C5, C9). Such emotional engagement makes reasoning and reflecting an enjoyable process, and may itself become a goal for using a feedback tool.

The core of our feedback model is a loop. Acting on **behavior choices** leads to changes in **state**. When the feedback tool is used for ongoing monitoring and reflection, the user continually **reasons** about their **state** to gain new **insights**, leading to new **behavior choices** and revised **personal goals**. Meanwhile, the feedback loop is embedded in a personal context, reflecting how feedback tool use varies with individual differences. Due to individual differences (e.g., physical conditions, domain knowledge, data analysis skills), people's monitoring measures, goals, reasoning strategies, and influences on behaviors vary.

### 5.4 Effects of the On-Calendar Visualization

In this section we discuss the effects of the on-calendar visualization within the feedback model above.

**Revealing state**: Participants reported that the on-calendar visualization was good for showing overall trends, consistent repeated patterns and peak values. *"This is great…when I'm at work it's pretty clear peaks for my morning walk, my lunch time and my home break whereas with the kids it's just sort of this little…and when I'm on my own there is peaks in intensity"* (V3). It could also be perceived simply with a glance (with line graphs): *"this is all information I am just gleaning from glance and I like that a lot."* (V2). On the other hand, there were some limitations of the visualization to present information about state. The tool made it difficult to monitor specific measures in detail since values were not labeled on the graph (e.g., total daily steps). In addition, although step data reflected exercise generally, participants found it challenging to segment different activities: *"It would be really nice if somewhere on here it would show when I played squash and then I could count the days since I had last played and that kind of thing"* (V2).

**Reasoning**: A real strength of the approach was that participants could easily relate data to calendar activities to explain data patterns, especially with the week view. We coded the instances when participants were not sure or could not figure out the reason for a data pattern. In the Control group this frequency dropped from 23 (first interview) to 15 (last interview), while in the Visualization group it dropped further, from 29 to 8, suggesting that the calendar visualization was more helpful for reasoning than the baseline tools.

Participants were excited to tell us their new insights during their use. V1 mostly put work events on her calendar. She noticed she was actually more active when working in the office than working at home. V9 recalled a concert experience (an event on his calendar). He was surprised to see (via the Fitbit data) that almost half of the time was for the intermission. V9 was able to reconstruct non-documented events by viewing the on-calendar visualization: *"I was just sitting around and then I went to yoga after that, so this probably indicates to say I went along to get some dinner and I ate the dinner and I walked to yoga. And then in the evening, I went for another walk"* (V9).

The calendar visualization also helped participants to identify and reason about patterns and anomalies. V1 and V8 identified the intense spikes from the running competition in the city. V8 noticed data spikes during an exam on her calendar, which she attributed to Fitbit capturing the hand movement. V10 identified days he commuted by bike or car.

Six participants also reported that the on-calendar visualization helped their awareness, e.g., *"I'm surprised that I'm actually*

17

*more active on the days where I have to go to work, as opposed to the days when I work from home. I thought I would be more active on those days [at home], but I take less breaks … but I find because I put like on a Google calendar, the day, hours that I'm working from home. So, I can remember it too and then it shows that I'm less active… that was surprise* " (V1) "*Whereas before it's kind of, you get so caught up in your life and your schedule and what you're doing, you might not even think that you haven't been out. But it's the awareness that helps.*" (V2)

**Behavior choice**: We found that the calendar visualization might not provide direct actionable insights to instruct users' behaviors; however, the influence on actions could be long term. Interestingly, V5 used the calendar visualization as a logging tool. She added calendar events expressly for the purpose of explaining the line graph patterns, e.g., "putting kids to bed", "dinner with family friends", etc. She reported that doing this helped her to recall and reason about her data patterns. The on-calendar application also helped participants to plan their exercises: *"Well I look at weeks and then I think in terms of, instead of a daily thing… the calendar has helped me focus on maybe a week or a month in advance and what I have to do."* (V6) C4 also mentioned when she put an exercise plan in her schedule, she was more likely to follow that plan. When she was introduced to the calendar in the end, she was exited to see it was what she needed, *"Yeah, so then I've the flexibility of determining where it fits and then if I could just come back and say did I run or whatever"*.

**Affective engagement**: The on-calendar application helped participants to recall their experiences, often evoking pleasant emotions. *"You know what is this? [on Eastern Sunday] we were hiding eggs in the midnight for the kids. [laugh]"* (V3) With the Fitbit data and events on one's calendar, they may reconstruct their life and go through the past. *"Yeah, I think so that will be cool, because then I could say like, the exact time when I met the person would be the time that I stop walking to talk to them…I do like the ability to look at my history and this is such a cool feature like, I will be sad when this study ends…."* (V9). The contextual framing of data (i.e., personal calendars) facilitated serendipitous exploration [21], eliciting emotion in reminiscence.

In contrast, emotion associated with feedback use for participants in the control group was limited to summary data: participants mentioned an emotional response to the smiley face that represented meeting their daily goals, weekly summary of weekly progress, or total steps in social challenges. However, we did not observe similar emotional mentions when they were looking at their raw Fitbit data or recalling related events.

### 5.5 Context for Reasoning

When we asked participants to investigate their data, they usually referred to contextual information. We coded all events when participants were trying to use contextual information to reason. The most frequently used types are shown in Table 1.

Most of the frequently used information could be found on the participants' personal calendars. All participants in the Visualization group reported that they liked having their Fitbit data and life events aligned together on a calendar. It was easy to access and also provided contextual information. Participants in both groups usually spontaneously brought up or referred to their personal calendar (17/19 participants). This observation confirmed our intuition that personal calendar events could provide relevant context for reasoning about temporal fitness data. Moreover, the on-calendar visualization makes this contextual information easier to access. The control group had to bring up their calendar as a separate application or on a different device.

In addition, the timeline of a day provided general context, for example the time to get up, to run for bus, to jog during lunch and to exercise in the evening. Especially with the week view, participants could glance at routines across the week. *"I can see that I was active only in the middle of the day and without knowing what the numbers are just comparing this day or this day, I know that like let's say this was roughly five or six p.m., I went straight home, but this is all information I'm just gleaning from glancing and I like that a lot."* (V2) C9 had even implemented a similar system (non-digital) 15 years ago. She logged exercise on a paper based wall calendar, and later put it into a spreadsheet that could be visualized using charts.

Table 1. Frequencies of using contextual information for reasoning (normalized as frequency per participant).

| Context Information | Control | Visualization |
| --- | --- | --- |
| Schedules and holidays | 2.6 | 3.1 |
| Family events | 2.4 | 1.0 |
| Social activities | 1.4 | 1.6 |
| Location | 0.7 | 1.7 |
| Life routines not on the calendar | 0.7 | 0.9 |

Meanwhile, other relevant context information could not be found on calendars. Data granularity on activity level was neither available in the calendar visualization nor in the Fitbit application. Even with the time-varying step data, spikes were still not informative. *"It's hard to make some days like this day… kind of hard for me to tell. Sometimes even if it's like kind of spikes, [it is still hard to tell]."* (V8) Participants reported that it would be helpful to identify and then compare different activities (V2, V9, C4, C7). For example, C7 reported that she usually ran on a treadmill that could tell her the distance, which enabled her to compare with the previous runs and see if she improved. *"I want to be able to look over time and say even though the steps go up and down, but my run has got longer or faster"* (C4). One challenge of the on-calendar visualization was that participants could not compare with baselines (e.g., daily goals, historical average or statistics representing performance in the population). V8, who was trying to investigate her sleep problem, reported that she expected to know how her data statistically fit in a larger population. In addition, participants felt they lacked domain knowledge to understand the measures (V1, V3, V6, V9, C1, C3, C4), for example how calories were calculated, what active minutes meant, and what was meant by heart rate zones.

### 5.6 Encouraging Ongoing Use

A feedback tool might aim to support ongoing tracking. However, ongoing use does not mean infinite use. We define ongoing use as the long-term adoption towards reaching one's personal goals. After reaching those goals, one's curiosity or interest may drop, or it might be possible to maintain consistent status without assistance from feedback tools. However, we would like to prevent premature discontinuation of tool use due to other reasons. We explore this matter through our feedback model (Figure 4). Any factor that might prevent the loop moving forward is likely to discourage further use of a feedback tool.

The first inhibitor to ongoing use is when the representation of state cannot reflect one's goals. V8 and V2 reported that steps were lower priority for them. Instead, they were more interested in sleep quality and records of playing squash respectively; our current calendar visualization did not support showing these data. In addition, how much effort one needs to dedicate to accessing and managing the application matters. We found that people who had greater access to their phone used feedback tools primarily on the phone, while those who spent more time working with computers used feedback primarily on their browser.

Second, a large gap between the state and one's personal goal can lead to frustration. V5 reported that she barely used the feedback applications for a few weeks, because she felt she was much less active than before and she did not want to see the trends going down. If one could not meet one's goal for a long time, frustration might make her/him drop the tool in the end.

Lack of support for reasoning might also prevent ongoing use. Domain knowledge might be required to interpret the data, or one may pick an inappropriate baseline (e.g., C4 found it was not realistic to compare herself with friends who were way more active than her.) This may limit meaningful insights and lead to feelings of powerlessness. Meanwhile, the common design strategy of representing the final results (e.g., status to represent active or inactive, or a pre-set heart rate zone) may be inadequate; this "black box" output can make people feel less involved, preventing ongoing use. Without knowing how calories were calculated, C4 chose to skip calorie-related features. V2 could not make sense how heart zone correlated with steps, so he had to search online resources for more knowledge.

Lack of actionable insights for behavior change and planning could also break the loop. For example, V3 knew she should exercise with her stationary bike, but did not know where to start, leaving the bike in the basement for hanging clothes.

Meanwhile, emotional engagement is an interesting factor in ongoing use. In our experience, interest and engagement in using feedback tools like Fitbit wanes over time; there is an initial period of novelty, followed by routine use, and an eventual drop-off. Our current study design was unable to assess whether deeper emotional engagement might encourage ongoing use, but this is an interesting topic to investigate in the future. This topic is likely tightly interwoven with social engagement, a strong motivator for behavior change. One may invite a friend to exercise together (V8, C4), or hire a trainer as commitment (V5). From the interviews, we found that people sometimes kept using a feedback tool just because her/his friends stayed with the same one (C9).

## 6 DISCUSSION

Here we reflect on our design approach and our feedback model with respect to related models from the literature. We also discuss the limitations of our research.

### 6.1 Reflection on the on-calendar visualization

Both groups of participants were very positive about integrating Fitbit data into a digital calendar; they found it easy to access and understand. The study also confirmed that personal digital calendars could provide rich contextual information for people to reason about their fitness data, and the on-calendar visualization made this information easy to access. Although digital calendars will not be able to capture all the context relevant to reasoning about one's physical activities, the study showed that the context they do provide could be helpful.

Our application was used in some ways we did not expect. One example was for logging: participant V5 added calendar events as a log to help recall and explain the Fitbit data at a later time. Interestingly, it also allowed participants to reflect on their past events and experiences; fitness data provided context for understanding past calendar events and activities not documented on the calendar at all. Similarly, people used it to plan fitness exercises that could fit nicely in their schedules. The mapping between scheduled events on one's calendar and the integrated fitness data could inform the user about their actual performance during the fitness sessions and could make them more accountable to follow their fitness plan. It could also support reminiscing. With easily accessed contextual information, people could re-experience affective responses associated with special moments while recalling the past. Some of these unanticipated uses turned out to be the most valuable attributes for some of our participants.

### 6.2 Feedback Model and Related Models

Our model (Figure 4) characterizes the role that feedback tools can play in evoking behavior change. While this model emerged from our qualitative content analysis, it does bear some resemblance to other models in the literature.

Feedback designs are usually connected with behavior change models. A variety of behavior models have been studied in practice [13], but most of these focus on how to affect people's motivation and attitude, and consequently influence behavior choices. However, the process of behavior change involves long-term learning and is on-going. Our interest with the feedback model is to explore how information design could facilitate this long-term process, for example, by making information more accessible and comprehensible. Although understanding one's data does not necessarily result in behavior change, we believe the role of feedback tools should also engage people in thinking and reflecting on information they receive; this may help people to set realistic and attainable goals, engaging them in the process.

As such, the feedback model enables designers to reflect on a non-persuasive approach. Tools that facilitate the reasoning process rather than enforcing behavior change might be one step forward towards encouraging people to change their behaviour on their own in a long-lasting way. Such tools need to present information that can be accessed easily and that reflect one's goal appropriately. Another implication is design for emotional engagement. Rather than being task-oriented, designers could consider designing for serendipitous information exploration.

In relation to more general models, the Technology Acceptance Model [2] is a well-known model of system adoption; however, it is not specific to feedback tools and does not consider the influence of technology on behaviors outside of tool use itself. A closely related but more specific model is the Promoter-Inhibitor Motivation Model (PIMM) [19]. PIMM models factors that promote and inhibit use of casual visualizations that people encounter in everyday life. However, our feedback model captures a specific case: on-going use of a feedback visualization to learn about and influence personal behavior. As such, while many of the influencing factors from PIMM still apply to our context, PIMM does not capture the role of insight in effecting behavior change, nor the subsequent (circular) effects on goals and motivations for using a feedback tool.

Our model is most certainly constrained by the scale and nature of our study. However, we believe it provides a starting point for thinking about how design characteristics might influence feedback tool use, including the likelihood of ongoing adoption and behavior change. Nonetheless, we fully expect that it will be revised with more input in future work.

### 6.3 Limitations of the Study

Since the application was independent (e.g., not implemented within google calendar or iCal) and had limited features (e.g., no custom coloring of calendar events), its use might have been constrained. At least one participant reported trying to manage both Google Calendar and our application. Participants expected that the data could be displayed in their own usual calendar.

In the field study, we recruited existing Fitbit users in hopes that they would use feedback tools on a regular basis. These people had previous experience using Fitbit's feedback tools, and may therefore react differently to our on-calendar visualizations than a more general population. Additionally, we did not control for use of Fitbit's feedback tools; it would have been difficult to constrain or track people's use of the Fitbit application except through unreliable self-reports.

In addition, our study had a small scale and only focused on physical activity. While we anticipate that on-calendar visualizations could be used to display any sort of personal quantitative feedback data (e.g., heart rate, blood pressure, resource use), future research is needed to understand user needs for other application domains. Meanwhile, the visualization layout was designed for desktop use, so future investigations could examine a version customized for mobile devices.

Our study is a starting point for exploring how to integrate personal data within a digital calendar. It suggests that designers may wish to further experiment with this and other non-persuasive approaches, and also give attention to engaging on-going use. Making contextual information easily accessible and blending feedback into currently used tools are promising design directions.

## 7 CONCLUSION

Providing context for reasoning and engaging long-term use are current challenges for design of feedback technologies. Towards these ends, we proposed a design approach of embedding personal quantitative data within a personal digital calendar. Our field study showed that this mash-up approach can provide easy access and offer contexual information to support reasoning about fitness data. We developed a model of the feedback process that extends existing technology adoption models. It also explains the role of feedback tools, thereby providing a structure for reasoning about feedback tool design choices and evaluation mechanisms. We hope this study will help designers reflect on non-persuasive approaches to feedback tool design.

## 8 ACKNOWLEDGMENTS

We thank our field study participants for their long-term commitment. This research was funded by NSERC and GRAND.## REFERENCES

[1] A. Andrew, G. Borriello, and J. Fogarty, "Toward a Systematic Understanding of Suggestion Tactics in Persuasive Technologies," in *Persuasive Technology*, Y. de Kort, W. IJsselsteijn, C. Midden, B. Eggen, and B. J. Fogg, Eds. Springer Berlin Heidelberg, 2007, pp. 259–270.

[2] A. Bajaj and S. R. Nidumolu, "A feedback model to understand information system usage," *Inf. Manage.*, vol. 33, no. 4, pp. 213–224, Mar. 1998.

[3] M. Barreto, E. Karapanos, and N. Nunes, "Why Don't Families Get Along with Eco-feedback Technologies?: A Longitudinal Inquiry," in *Proc. Biannual Conference of the Italian Chapter of SIGCHI*, USA, 2013, pp. 16:1–16:4.

[4] L. Bartram, "Design Challenges and Opportunities for Eco-Feedback in the Home," *IEEE Comput. Graph. Appl.*, vol. 35, no. 4, pp. 52–62, Jul. 2015.

[5] S. Consolvo, K. Everitt, I. Smith, and J. A. Landay, "Design requirements for technologies that encourage physical activity," in *Proc. SIGCHI Conference on Human Factors in Computing Systems*, USA, 2006, pp. 457–466.

[6] S. Consolvo, D. W. McDonald, T. Toscos, M. Y. Chen, J. Froehlich, B. Harrison, P. Klasnja, A. LaMarca, L. LeGrand, R. Libby, I. Smith, and J. A. Landay, "Activity Sensing in the Wild: A Field Trial of Ubifit Garden," in *Proc. SIGCHI Conference on Human Factors in Computing Systems*, USA, 2008, pp. 1797–1806.

[7] E. Costanza, S. D. Ramchurn, and N. R. Jennings, "Understanding domestic energy consumption through interactive visualisation: a field study," in *Proc. 2012 ACM Conference on Ubiquitous Computing*, USA, 2012, pp. 216–225.

[8] T. Erickson, M. Podlaseck, S. Sahu, J. D. Dai, T. Chao, and M. Naphade, "The dubuque water portal: evaluation of the uptake, use and impact of residential water consumption feedback," in *Proc. SIGCHI Conference on Human Factors in Computing Systems*, USA, 2012, pp. 675–684.

[9] T. Fritz, E. M. Huang, G. C. Murphy, and T. Zimmermann, "Persuasive Technology in the Real World: A Study of Long-term Use of Activity Sensing Devices for Fitness," in *Proc. SIGCHI Conference on Human Factors in Computing Systems*, USA, 2014, pp. 487–496.

[10] J. Froehlich, L. Findlater, and J. Landay, "The design of eco-feedback technology," in *Proc. SIGCHI Conference on Human Factors in Computing Systems*, USA, 2010, pp. 1999–2008.

[11] J. Froehlich, L. Findlater, M. Ostergren, S. Ramanathan, J. Peterson, I. Wragg, E. Larson, F. Fu, M. Bai, S. Patel, and J. A. Landay, "The design and evaluation of prototype eco-feedback displays for fixture-level water usage data," in *Proc. the SIGCHI Conference on Human Factors in Computing Systems*, USA, 2012, pp. 2367–2376.

[12] M. Haller, C. Richter, P. Brandl, S. Gross, G. Schossleitner, A. Schrempf, H. Nii, M. Sugimoto, and M. Inami, "Finding the Right Way for Interrupting People Improving Their Sitting Posture," in *Human-Computer Interaction – INTERACT 2011*, vol. 6947, P. Campos, N. Graham, J. Jorge, N. Nunes, P. Palanque, and M. Winckler, Eds. Springer Berlin / Heidelberg, 2011, pp. 1–17.

[13] H. A. He, S. Greenberg, and E. M. Huang, "One size does not fit all: applying the transtheoretical model to energy feedback technology design," in *Proc. the SIGCHI Conference on Human Factors in Computing Systems*, USA, 2010, pp. 927–936.

[14] D. Huang, M. Tory, B. Adriel Aseniero, L. Bartram, S. Bateman, S. Carpendale, A. Tang, and R. Woodbury, "Personal Visualization and Personal Visual Analytics," *IEEE Trans. Vis. Comput. Graph.*, vol. 21, no. 3, pp. 420–433, Mar. 2015.

[15] P. Klasnja, S. Consolvo, and W. Pratt, "How to evaluate technologies for health behavior change in HCI research," in *Proc. the SIGCHI Conference on Human Factors in Computing Systems*, USA, 2011, pp. 3063–3072.

[16] J. Lin, L. Mamykina, S. Lindtner, G. Delajoux, and H. Strub, "Fish'n'Steps: Encouraging Physical Activity with an Interactive Computer Game," in *UbiComp 2006: Ubiquitous Computing*, vol. 4206, P. Dourish and A. Friday, Eds. Springer Berlin / Heidelberg, 2006, pp. 261–278.

[17] M. B. Miles and M. Huberman, *Qualitative Data Analysis*, 3rd edition. Thousand Oaks, Calirorinia: Sage Publications, 2013.

[18] C. Neustaedter, L. Bartram, and A. Mah, "Everyday activities and energy consumption: how families understand the relationship," in *Proc. the SIGCHI Conference on Human Factors in Computing Systems*, USA, 2013, pp. 1183–1192.

[19] D. W. Sprague and M. Tory, "Exploring How and Why People Use Visualizations in Casual Contexts: Modeling User Goals and Regulated Motivations," *Information Visualization*, vol. 11, no. 2, pp. 106–123, Apr. 2012.

[20] Y. Strengers, "Peak electricity demand and social practice theories: Reframing the role of change agents in the energy sector," *Energy Policy*, vol. 44, pp. 226–234, May 2012.

[21] A. Thudt, D. Baur, S. Huron, and S. Carpendale, "Visual Mementos: Reflecting Memories with Personal Data," *IEEE Transactions on Visualization and Computer Graphics*, vol. 22, no. 1, pp. 369–378, Jan. 2016.

[22] F. Yetim, "Critical Perspective on Persuasive Technology Reconsidered," in *Proc. SIGCHI Conference on Human Factors in Computing Systems*, USA, 2013, pp. 3327–3330.

[23] "International Physical Activity Questionnaire." [Online]. Available: https://sites.google.com/site/theipaq/home. [Accessed: 17-Sep-2015].

[24] "AQUAD." [Online]. Available: http://www.aquad.de/en/. [Accessed: 20-Sep-2015].20